\documentclass[useAMS, usenatbib]{mn2e}
\usepackage{graphicx,amsmath,color}

\begin{document}

 \title[Multi-frequency Synthesis]
 {Multi-frequency synthesis algorithm based on Generalized Maximum Entropy Method.
 Application to 0954+658}
 \author[A.T. Bajkova \& A.B. Pushkarev]
  {Anisa T.~Bajkova$^1$\thanks{E-mail: anisabajkova@rambler.ru} and Alexander B. Pushkarev$^{1,2}$\\
 $^1$Central (Pulkovo) Astronomical Observatory of RAS,
 Pulkovskoe Chaussee 65/1, Saint-Petersburg, 196140, Russia\\
 $^2$Crimean Astrophysical Observatory, 98409 Nauchny, Crimea, Ukraine\\
 }

\date{Accepted 2011 June 17. Received 2011 June 14; in original form
2011 March 18}

\pagerange{\pageref{firstpage}--\pageref{lastpage}} \pubyear{2011}
\def\LaTeX{L\kern-.36em\raise.3ex\hbox{a}\kern-.15em
    T\kern-.1667em\lower.7ex\hbox{E}\kern-.125emX}
\newtheorem{theorem}{Theorem}[section]
\label{firstpage}
\maketitle

\begin{abstract}
We propose the multi-frequency synthesis (MFS) algorithm with
spectral correction of frequency-dependent source brightness
distribution based on maximum entropy method. In order to take
into account the spectral terms of n-th order in the Taylor
expansion for the frequency-dependent brightness distribution, we
use a generalized form of the maximum entropy method suitable for
reconstruction of not only positive-definite functions, but also
sign-variable ones. The proposed algorithm is aimed at producing
both improved  total intensity image and two-dimensional spectral
index distribution over the source. We consider also the problem
of frequency-dependent variation of the radio core positions of
self-absorbed active galactic nuclei, which should be taken into
account in a correct multi-frequency synthesis. First, the
proposed MFS algorithm has been tested on simulated data and then
applied to four-frequency synthesis imaging of the radio source
0954+658 from VLBA observational data obtained
quasi-simultaneously at 5, 8, 15 and 22~GHz.
\end{abstract}


\section{Introduction}

At present, very-long-baseline interferometry (VLBI) is the most
powerful tool for studying the morphological structures as well as
kinematic, polarization, and spectral characteristics of active
galactic nuclei (AGN). It allows objects to be imaged with a very
high angular resolution reaching fractions of a milliarcsecond
(mas). One of the topical problems of VLBI mapping of AGN is
multi-frequency image synthesis. Our interest in this method is
mainly related to the peculiar geometry of the future high-orbit
ground-space radio interferometer {\it Radioastron}
\citep{kard97}, which is expected to provide an ultra-high
resolution (microarcseconds), but poor aperture filling
\citep{bajk05}.

Multi-frequency synthesis in VLBI suggests mapping AGN at several
frequencies simultaneously to improve the instrument aperture
filling. This is possible, because the interferometer baselines
are measured in wavelengths of the emission being received. The
problem of multi-frequency synthesis is complicated due to the
frequency-dependent brightness of a radio source. Hence, to avoid
undesirable artifacts in the reconstructed image, spectral
correction should be made at the stage of its deconvolution.

\cite{ccw90}, \cite{conw91}, \cite{sault94}, and \cite{sault99}
investigated the influence of spectral effects on the image and
developed the methods of their correction. These authors showed
that if narrow frequency bands, up to $\pm$12.5\% of the reference
frequency, are used, the effects of the spectral dependence of the
brightness of a radio source can usually be ignored for dynamic
ranges of less than 1000:1. When the spectral errors are above the
noise they can be recognized and removed to ensure required
dynamical range of images. The spectral errors can usually be
accounted for by parameterizing the image in terms of two unknowns
at each pixel: the intensity at some reference frequency and
spectral index if spectral variation of the source emission is
modelled by a power-law relationship. Thus, the use of MFS doubles
the number of unknowns but in case of MFS we have $n_f$ more data
of observations, where $n_f$ is number of frequencies. As
discussed in paper by \cite{conw91}, if $n_f$ greater than 2 the
MFS image remains better constrained than the single frequency
image.

The algorithm of linear spectral correction based on the CLEAN
method \citep{hogb74} and called ``double deconvolution''
\citep{ccw90} is the best-studied one. In this algorithm, the
``dirty'' image is first deconvolved with an ordinary ``dirty''
beam and the residual map is then deconvolved with the beam
responsible for the first-order spectral term. An improvement of
this method consisting in simultaneous reconstruction of the
sought-for image and the map of the spectral term was proposed by
\cite{sault94}. The vector relaxation algorithm developed by
\cite{llg06} may be considered as a generalized CLEAN
deconvolution method that can account for spectral terms of any
order. Description of the MFS technique in a common mathematical
framework is given in \cite{rau09} and \cite{rau10}.

An alternative deconvolution method also actively used in radio
astronomy is the maximum entropy method (MEM). MEM was first
proposed by \cite{fried72} and \cite{abl74} for the reconstruction
of images in optics and radio astronomy, respectively. Since then
the method has been developed in many studies \citep{skil84,
narnit86, wern76, coreva85, frbajk94} and implemented in a number
of software packages designed for image reconstruction (MEMSYS,
AIPS, etc.). A comparative analysis of CLEAN and MEM in VLBI is
given in \cite{cbb99}; essentially, these methods complement each
other. In particular, CLEAN is preferred for reconstructing images
of compact sources from relatively poor data, while MEM is more
suitable for imaging of extended sources from better-quality data.

A severe drawback of MEM compared to CLEAN, the bias of the
solution \citep{cbb99}, can easily be removed by generalizing the
method to enable the reconstruction of sign-variable functions
\citep{bajk92, frbajk94}. This generalized MEM also permits
difference imaging making it possible to substantially broaden the
dynamic range of maps of sources, including both compact and
extended, faint components \citep{bajk07,rast11}.

As shown by \cite{bajk08}, applying Shannon's maximum entropy
method allows a tangible progress to be achieved in solving the
problem of multi-frequency synthesis owing to the possibility of a
simple allowance for the spectral terms of any order. This, in
turn, allows the range of synthesized frequencies to be extended
significantly. However, the multi-frequency synthesis algorithms
based on both CLEAN and MEM deconvolution discussed here can be
directly applied only to those radio sources for which no
frequency-dependent image shift is observed. Otherwise, as shown
by \cite{crogab08}, an additional operation to align images at
different frequencies should be performed to obtain the proper
results in a multi-frequency data analysis.

In this paper, our goal is to deduce our multi-frequency synthesis
algorithm based on MEM and to show the importance of the procedure
for precorrecting the frequency-dependent image shift while
implementing multi-frequency synthesis.

The paper is structured as follows. The frequency dependence of
the image of a radio source is described in Section
\ref{s:freq_dep}. Frequency-dependent constraints on the
visibility function are derived in Section \ref{s:constr}. The
reconstruction method is given in Section \ref{s:meth}. Prior to
describing the MFS algorithm, we consider the maximum entropy
method and its generalized form in Subsections \ref{ss:mem} and
\ref{ss:gmem}, respectively. The proposed multi-frequency
synthesis algorithm with frequency correction is deduced in
Subsection \ref{ss:mfs}. We test our algorithm in a number of
model experiments in Section \ref{s:test}. Discussion of the
problem of aligning frequency-dependent images to properly
construct the spectral index distribution is given in Section
\ref{s:align}. And, finally, in Section \ref{s:proc}, we present
the results of applying the MFS algorithm proposed to
four-frequency VLBA data for BL~Lacertae object 0954+658.

\section{Frequency dependence of the AGN radio brightness}
\label{s:freq_dep}

The dependence of the intensity of a radio source on frequency
$\nu$ in the model of synchrotron radiation is given by
\citep{ccw90}
\begin{equation}
\label{eq:1}
I(\nu)=I(\nu_0)\left(\frac{\nu}{\nu_0}\right)^\alpha,
\end{equation}
where $I(\nu_0)$ is the intensity of the radiation at the
reference frequency $\nu_0$ and $\alpha$ is the spectral index. To
simplify the writing, we will hereafter set $I_0=I(\nu_0)$.

Retaining the first $Q$ terms in the Taylor expansion of
(\ref{eq:1}) at point $\nu_0$, we can write the following
approximate equality:
\begin{equation}
\label{eq:2}
I(\nu)\approx I_0+\sum_{q=1}^{Q-1}I_q
\left(\frac{\nu-\nu_0}{\nu_0}\right)^q,
\end{equation}
where
$$I_q=I_0\frac{\alpha(\alpha-1)\cdots[\alpha-(q-1)]}{q!}.$$

From (\ref{eq:2}), for each pixel $(k,l)$ of the source's
two-dimensional ($N\times N$) brightness distribution we have
\begin{equation}
\label{eq:3}
I(k,l)\approx I_0(k,l)+\sum_{q=1}^{Q-1}I_q(k,l)
\left(\frac{\nu-\nu_0}{\nu_0}\right)^q,
\end{equation}
where $k,l=1,\ldots,N$.

Thus, the derived brightness distribution over the source
(\ref{eq:3}) is the sum of the brightness distribution at the
reference frequency $\nu_0$ and the spectral terms with the
$q$th-order spectral map depending on the spectral index
distribution over the source as follows:
\begin{equation}
\label{eq:4}
I_q(k,l)=I_0(k,l)\frac{\alpha(k,l)\cdots[\alpha(k,l)-(q-1)]}{q!}.
\end{equation}
Of greatest interest is the first-order spectral map
$$
I_1(k,l)=I_0(k,l)\alpha(k,l),
$$
because the spectral index distribution over the source can be
estimated from that in the following way:
\begin{equation}
\label{eq:5}
\alpha(k,l)=I_1(k,l)/I_0(k,l).
\end{equation}

\section{Constraints on the visibility function}
\label{s:constr}

The complex visibility function is the Fourier transform of the
intensity distribution over the source that satisfies the spectral
dependence (\ref{eq:1}) at each pixel of the map $(k,l)$. Given
the finite number of terms in the Taylor expansion (\ref{eq:3}),
the constraints on the visibility function ${\bf V}$ can be
written as
\begin{eqnarray}
\label{eq:6}
 V_{u_\nu,v_\nu}= {\bf F}\{I(k,l)\}\times {\bf D}_{u_\nu,v_\nu}\\
\approx \sum_{q=0}^{Q-1} {\bf F}\left\{I_q(k,l)
\left(\frac{\nu-\nu_0}{\nu_0}\right)^q\right\}\times {\bf
D}_{u_\nu,v_\nu},\nonumber
\end{eqnarray}
where {\bf F}  denotes the Fourier transform and {\bf D}
represents the transfer function, which is the $\delta$-function
of $u$ and $v$ for each measurement of the visibility function;
different sets of $\delta$-functions correspond to different
frequencies $\nu$, as suggested by the indices of $u$ and $v$.

Let us rewrite (\ref{eq:6}) for the real and imaginary parts of
the visibility function
$V_{u_\nu,v_\nu}=A_{u_\nu,v_\nu}+jB_{u_\nu,v_\nu}$ by taking into
account the measurement errors as
\begin{equation}
\label{eq:7} \sum_{q=0}^{Q-1}\sum_{k,l} I_q(k,l)
a^{lm}_{u_\nu,v_\nu}\left(\frac{\nu-\nu_0}{\nu_0}\right)^q+\eta^{\rm
re}_{u_\nu,v_\nu}= A_{u_\nu,v_\nu},
\end{equation}
\begin{equation}
\label{eq:8} \sum_{q=0}^{Q-1}\sum_{k,l} I_q(k,l)
b^{lm}_{u_\nu,v_\nu}\left(\frac{\nu-\nu_0}{\nu_0}\right)^q+\eta^{\rm
im}_{u_\nu,v_\nu}= B_{u_\nu,v_\nu},
\end{equation}
where $a^{lm}_{u_\nu,v_\nu}$ and $b^{lm}_{u_\nu,v_\nu}$ are the
constant coefficients (cosines and sines) that correspond to the
Fourier transform, $\eta^{\rm re}_{u_\nu,v_\nu}$ and $\eta^{\rm
im}_{u_\nu,v_\nu}$  are the real and imaginary parts of the
instrumental additive noise distributed normally with a zero mean
and a known dispersion $\sigma_{u_\nu,v_\nu}$.

\section{Method}\label{s:meth}

\subsection{Maximum Entropy Method} \label{ss:mem}

MEM is one of a large class of non-linear informational methods
based on the optimization of a functional specified by some
informational criterion for the quality of the solution subject to
the constraints that flow from the data. In our case, maximizing
the Shannon entropy consists in finding the maximum of the
functional
\begin{equation}
\label{eq:9} E = -\int x(t) \ln(x(t))\>\rmn{d}t,
\end{equation}
where $x(t)$ is the desired distribution.

Since imaging in VLBI implies dealing with digital data, we
present a discrete formulation of the optimization. Let a map of
an object with a finite carrier be sampled in accordance with
Kotelnikov-Nyquist theorem \citep{opp99} and have a size of
$N\times N$ pixels. We denote the discrete measurements of the
desired distribution by
$$
x_{kl},\quad k,l=1,\ldots,N-1.
$$
We denote the known measurements of the two-dimensional Fourier
spectrum of the object, which represent the visibility data, in
accordance with the Van Cittert-Cernike theorem, as follows,
separating the real, $A_m$, and imaginary, $B_m$, parts:
$$
V_m = A_m + jB_m,\quad m=1,\ldots,M,
$$
where $M$ is the number of known measurements and $m$ is the
number of the current measurement with coordinates $(u_m,v_m)$ in
the {\it uv}-plane, not necessarily located at nodes of the
coordinate grid. This last circumstance means that there is no
problem with pixelization of the data in the frequency domain,
which represents a certain technical advantage of this method over
other methods and appreciably enhances the accuracy of the
reconstruction.

The practical MEM algorithm we applied, taking into account the
errors in the data \citep{bajk93}, implies the solution of the
conditional optimization problem
\begin{equation}
\label{eq:10} \min \sum_k\sum_l
x_{kl}\ln(x_{kl})+\frac{1}{2}\sum_m\frac{(\eta_m^{\rm
re})^2+(\eta_m^{\rm im})^2}{\sigma_m^2},
\end{equation}
\begin{equation}
\label{eq:11} \sum_k\sum_l x_{kl}a_{kl}^m+\eta_m^{\rm re}=A_m,
\end{equation}
\begin{equation}
\label{eq:12} \sum_k\sum_l x_{kl}b_{kl}^m+\eta_m^{\rm im}=B_m,
\end{equation}
\begin{equation}
\label{eq:13} x_{kl}\ge 0.
\end{equation}

As we can see from (\ref{eq:10}), the optimized functional has two
parts: a Shannon entropy functional and a functional that is an
estimate of the difference between the reconstructed spectrum and
the measured data according to the $\chi^2$ criterion. This latter
functional can be considered an additional regulating, or
stabilizing, term acting to provide a further regularization of
the MEM solution. The influence of this additional term on the
resolution of the reconstruction algorithm must be kept in mind.

Equations (\ref{eq:11}), (\ref{eq:12}) represent linear
constraints on the unknown images $x_{kl}$  as well as noise terms
$\eta_m^{\rm re}$ and $\eta_m^{\rm im}$. The non-negativity
constraint (\ref{eq:13}) on the image can be omitted in this case
due to the nature of the entropy solution, which is purely
positive. If the total flux of the source $F_0$ is known, this
automatically leads to the normalization of the solution:
$$
\sum_k\sum_l x_{kl} = F_0.
$$
The numerical algorithm for solution (\ref{eq:10})--(\ref{eq:12}),
treated as a non-linear optimization problem based on the method
of Lagrange multipliers, is considered in detail in \cite{bajk07}.
Here, we present only the solution:
\begin{equation}
\label{eq:14}
x_{kl}=\exp(-\sum_m(\alpha_m a_{kl}^m+ \beta_m
b_{kl}^m)-1),
\end{equation}

$$
\eta_m^{\rm re} = \sigma_m^2 \alpha_m, ~~~~\eta_m^{\rm im} =
\sigma_m^2 \beta_m, $$
expressed in terms of the Lagrange
multipliers (dual variables) $\alpha_m$ and $\beta_m$, through
which the constraints (\ref{eq:11}) and (\ref{eq:12}),
respectively, enter the Lagrange functional.

As we can see from (\ref{eq:14}), the standard MEM image is
evidently positive. It can be shown that the MEM Hesse matrices
everywhere are positive-definite, so that the entropy functional
is convex and the solution is global. Various gradient methods can
be used to search for the extrema of the corresponding dual
functional. We use a coordinate-descent method.

\subsection{Generalized Maximum Entropy Method} \label{ss:gmem}

The GMEM was designed for the reconstruction of sign-variable and
complex functions \citep{bajk92, frbajk94, bajk07}. For the GMEM,
dealing with sign-variable real distributions, the Shannon-entropy
functional has the form
$$
E= -\int \{x^+(t) \ln([a x^+(t)] + x^-(t) \ln[a
x^-(t)]\}\>\rmn{d}t, $$
where $x^+(t)\ge 0$ and $x^-(t)\ge 0$ are
the positive and negative components of the sought-for image
$x(t)$, i.e. the equation $x(t) = x^+(t) - x^-(t)$ holds; $a
> 0$ is a parameter responsible for the accuracy of the separation
of the negative and positive components of the solution $x(t)$,
and therefore critical for the resulting image fidelity. As it was
shown by \cite{bajk07} solutions for $x^+(t)$ and $x^-(t)$
obtained with the Lagrange optimization method are connected by
the expression
$$
x^+(t) \cdot x^-(t) = \exp(-2-2\ln a) = K(a),
$$
which depends only on the parameter $a$.

This parameter is responsible for dividing the positive and
negative parts of the solution: the larger $a$ allows the more
accurate discrimination (since $K(a) \to 0$ as $a \to \infty$). On
the other hand, the value of $a$ is constrained by computational
limitations. The main constraint comes from the $\chi^2$ term in
the optimized functional, which depends on the data errors. The
larger a standard deviation is, the higher value of $a$ could be
set. If data are very accurate, a lower value of $a$ is needed. In
practice, $a$ is chosen empirically. In our case we had to
compromise between data errors, which determine resolution of the
final MEM solution, and a need to divide the positive and negative
parts of the solution as accurately as possible. It is fair to say
that given fixed errors in the data, a maximum possible chosen
value of $a$ provides us with the best possible resolution of the
MEM-solution. In this work we used $a = 10^6$.

\subsection{GMEM-based MFS algorithm} \label{ss:mfs}

In this case, the distributions $I_q(k,l)$, $q=0,\ldots,Q-1;$
$k,l=1,\ldots,N$,  and the measurement errors of the visibility
function $\eta^{\rm re}_{u_\nu,v_\nu}$, $\eta^{\rm
im}_{u_\nu,v_\nu}$ are unknown. Note that although the brightness
distribution over the source is described by  a non-negative
function, the spectral maps of arbitrary order (\ref{eq:4}) can
generally take both positive and negative values because the
spectral index distribution over the source is an alternating one.

By setting, in accordance with the approach described above:
$$
I_q(k,l)=I_q^+(k,l)-I_q^-(k,l),~~~q=1,\ldots,Q-1,
$$
we obtain the following entropic functional to be minimized:
\begin{eqnarray}
\label{eq:23}
{\bf E}=\sum_{k,l} I_0(k,l)\ln [a I_0(k,l)]\\
+\sum_{q=1}^{Q-1} \sum_{k,l} \{I_q^+(k,l)\ln [a
I_q^+(k,l)]
+I_q^-(k,l)\ln [a I_q^-(k,l)]\}\nonumber\\
+\frac{1}{2}\sum_{u_\nu,v_\nu}\frac{(\eta^{\rm
re}_{u_\nu,v_\nu})^2+(\eta^{\rm
im}_{u_\nu,v_\nu})^2}{\sigma^2_{u_\nu,v_\nu}},\nonumber
\end{eqnarray}
$$
I_0(k,l)\ge 0,~~~~~I_q^+(k,l)\ge 0,~~~~~I_q^-(k,l)\ge 0.
$$

The linear constraints (\ref{eq:7}) and (\ref{eq:8}) on the
measured visibility function will be rewritten accordingly:
\begin{eqnarray}
\label{eq:24}
\sum_{k,l} I_0(k,l) a^{kl}_{u_\nu,v_\nu}\\
+\sum_{q=1}^{Q-1}\sum_{k,l} [I_q^+(k,l)-I_q^-(k,l)]
a^{lm}_{u_\nu,v_\nu}\nonumber\\
\times\left(\frac{\nu-\nu_0}{\nu_0}\right)^q+\eta^{\rm
re}_{u_\nu,v_\nu}= A_{u_\nu,v_\nu},\nonumber
\end{eqnarray}
\begin{eqnarray}
\label{eq:25} \sum_{k,l} I_0(k,l)
b^{lm}_{u_\nu,v_\nu}\\
+\sum_{q=1}^{Q-1}\sum_{k,l} [I_q^+(k,l)-I_q^-(k,l)]
b^{lm}_{u_\nu,v_\nu}\nonumber\\
\times\left(\frac{\nu-\nu_0}{\nu_0}\right)^q+\eta^{\rm
im}_{u_\nu,v_\nu}= B_{u_\nu,v_\nu}.\nonumber
\end{eqnarray}

Minimizing the functional (\ref{eq:23}) with constraints
(\ref{eq:24})--(\ref{eq:25}) constitutes the essence of the
MEM-based multi-frequency synthesis algorithm that seeks the
solution for all unknown
$I_0(k,l),~I_q^{+(-)}(k,l),~q=1,\ldots,Q-1,~k,l=1,\ldots,N$ and
$\eta^{\rm re}_{u_\nu,v_\nu},~\eta^{\rm im}_{u_\nu,v_\nu}$. A
detailed algorithm for numerical implementation of the proposed
multi-frequency synthesis method is given in \cite{bajk08}.

\section{Testing the method. Simulation results}
\label{s:test}

\begin{figure}
\begin{center}
\includegraphics[width = 0.45\textwidth] {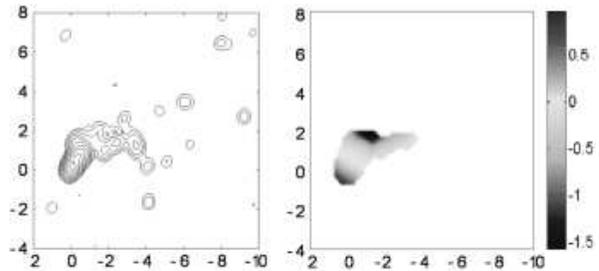}
\caption{Intensity map for the model radio source at 13.6~GHz
(left) and model spectral index map (right). Relative right
ascension and declination are given in mas along the horizontal
and vertical axes, respectively. Contour levels: 0.2, 0.4, 0.8,
1.6, 3.2, 6.4, 12.8, 25.6, 51.2, 90\% of the peak flux density of
0.196~Jy/pixel.} \label{fig:mod13}
\end{center}
\end{figure}

Here we present the results of testing our MFS deconvolution
algorithm on the example of four-frequency synthesis using
simulated VLBI data at four frequencies of 5, 8, 15 and 22~GHz. As
a reference frequency in the MFS algorithm, we adopted the central
frequency equal to 13.6 GHz. A model source map at this frequency
and a model spectral index distribution over the source are
presented in Fig.~\ref{fig:mod13}, left and right respectively.

\begin{table}
\caption{Parameters of model source images at different
frequencies.} \label{tbl:t1}
\begin{center}

\begin{tabular}{cccc}
\hline
     Frequency &  $S_{\rm tot}$ &    $S_{\rm peak}$ &   Entropy \\
         (GHz) &           (Jy) &        (Jy/pixel) &           \\
\hline
\phantom{1}5.0 &           5.17 &             0.187 &  $-$17.8 \\
\phantom{1}8.4 &           4.76 &             0.192 &  $-$16.2 \\
          13.6 &           4.64 &             0.196 &  $-$15.5 \\
          15.3 &           4.63 &             0.197 &  $-$15.4 \\
          22.2 &           4.65 &             0.201 &  $-$15.3 \\
\hline
\end{tabular}

\end{center}
\end{table}

As one can see, the source shows a structure on a milliarcsecond
angular scale consisting of a bright compact core and a one-sided
jet. Note that the model map contains also a number of weak
small-scale details scattered around the main structure. Such a
complication of the source structure was made in order to test the
ultimate capabilities of the MFS algorithm. Note also that the
structure of the model source was built similar to the structure
of the radio source 0954+658 which will be considered in
Section~\ref{s:proc}.

\begin{figure}
\begin{center}
\includegraphics[width = 40mm, height = 40mm]   {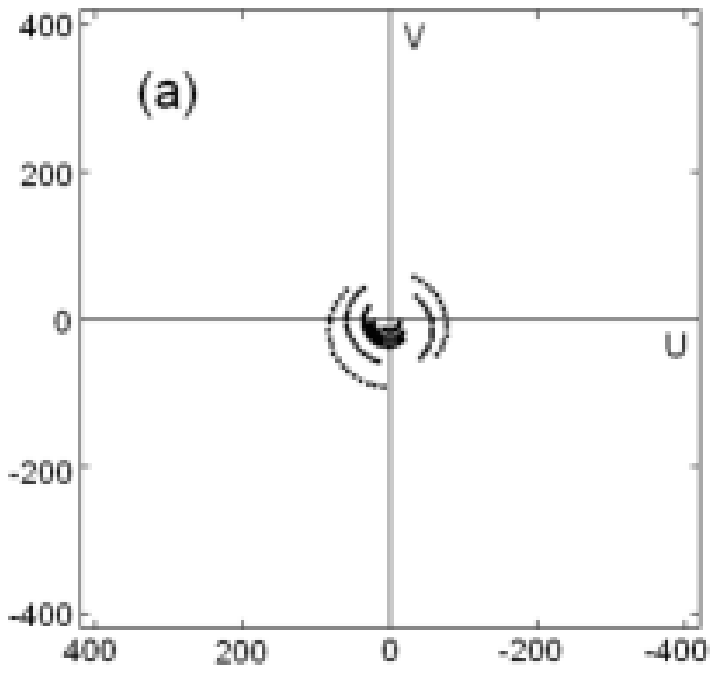}
\includegraphics[width = 40mm, height = 40mm]   {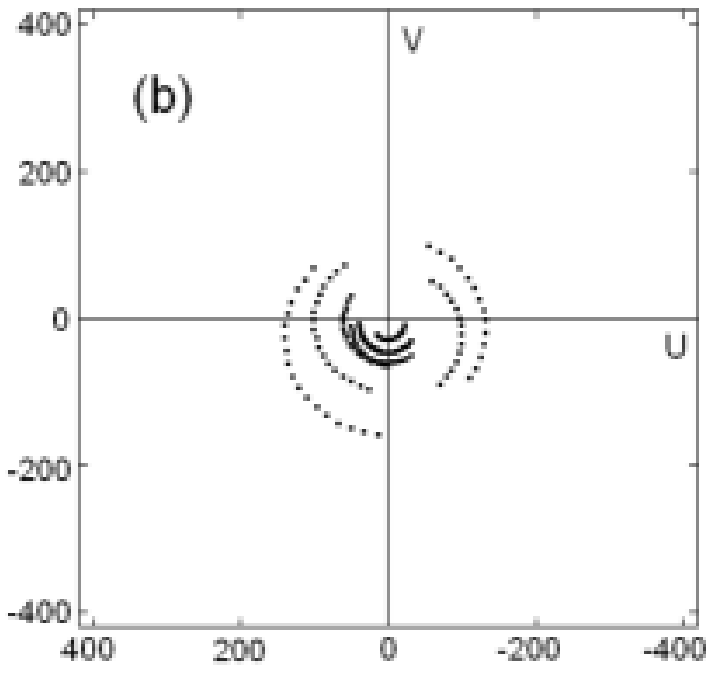}
\includegraphics[width = 40mm, height = 40mm]   {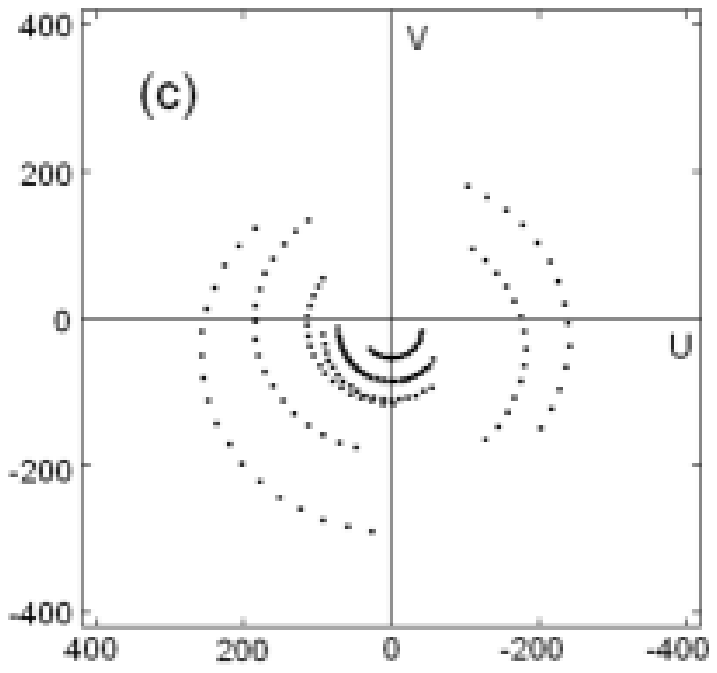}
\includegraphics[width = 40mm, height = 40mm]   {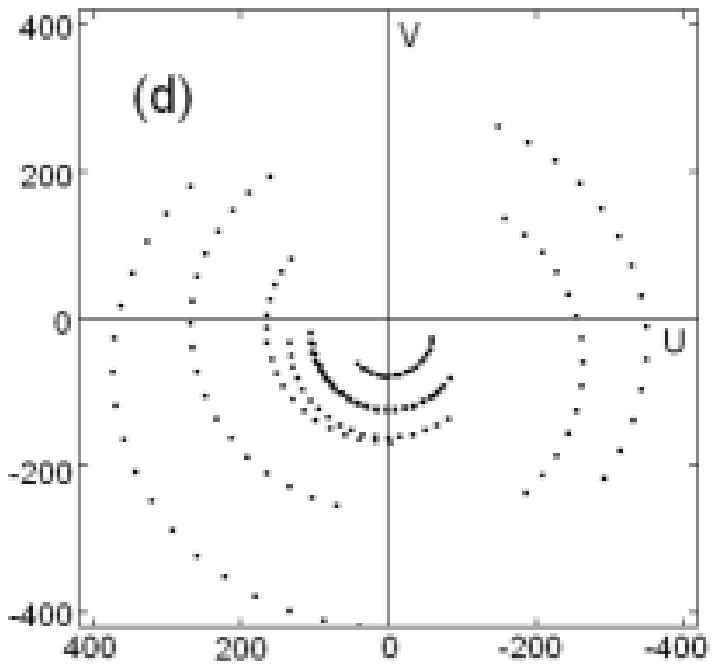}
\caption{Coverages of the {\it uv}-plane of the simulated
10-baseline interferometer related to data formed at 5 (a), 8 (b),
15 (c) and 22 (d) GHz. The axes are in units of millions of
wavelengths.} \label{fig:moduv}
\end{center}
\end{figure}

The {\it uv}-plane coverages related to a model interferometer
consisting of 10 baselines of the VLBA array (BR-LA, BR-MK, BR-NL,
BR-PT, BR-OV, BR-SC, BR-KP, BR-HN, BR-FD, LA-MK)  and source with
a declination $\delta=70\degr$ for each of ``observation''
frequencies are shown in Fig.~\ref{fig:moduv}. Note that the
choice of baselines was not principal for our task and was made in
an arbitrary way. It was assumed that a duration of the
observations is nine hours and visibility data are formed every 30
minutes.  As it can be seen, the {\it uv}-plane at different
frequencies has the same topology but scaled accordingly. As far
as interferometer baselines are measured in wavelengths, $u$- and
$v$- visibility coordinates are proportional to an observation
frequency.

\begin{figure}
\begin{center}
\includegraphics[width = 40mm, height = 40mm]   {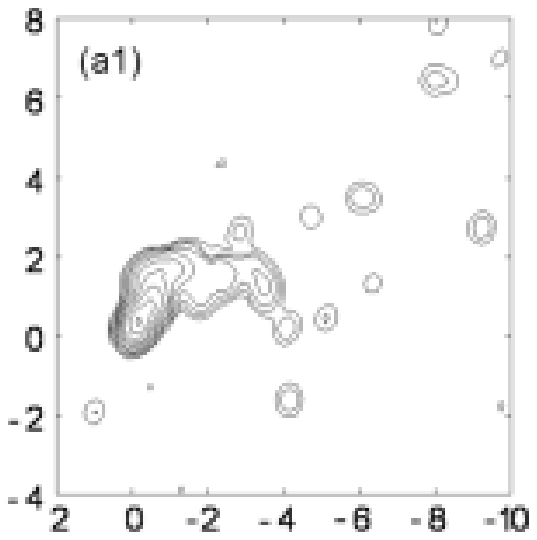}
\includegraphics[width = 40mm, height = 40mm]   {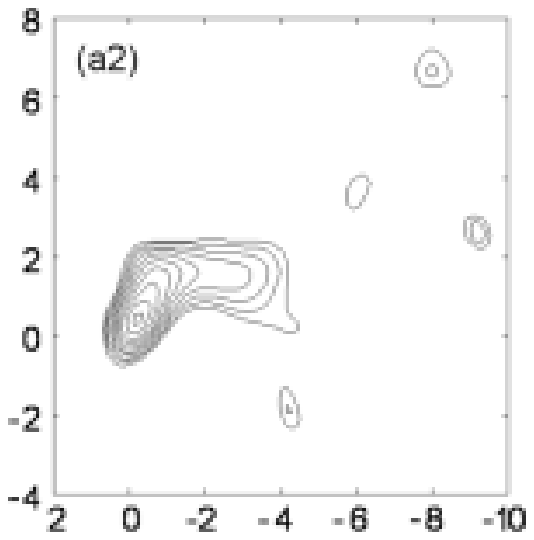}
\includegraphics[width = 40mm, height = 40mm]   {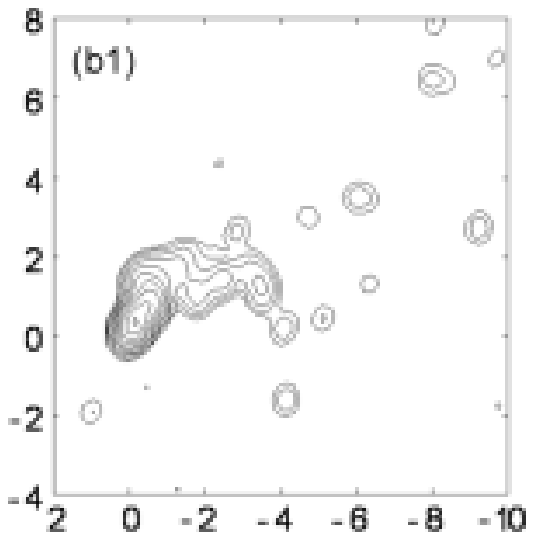}
\includegraphics[width = 40mm, height = 40mm]   {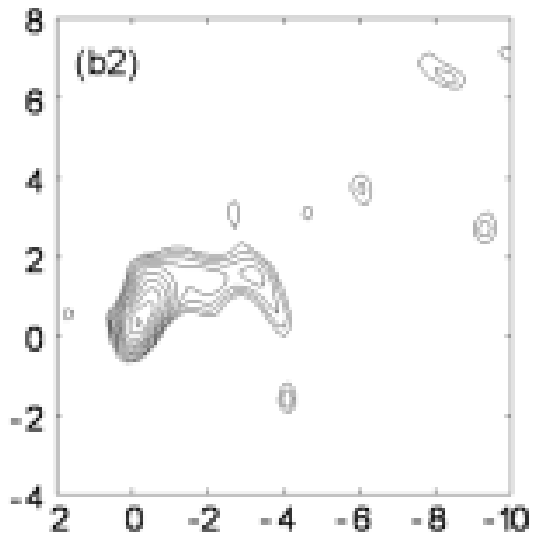}
\includegraphics[width = 40mm, height = 40mm]   {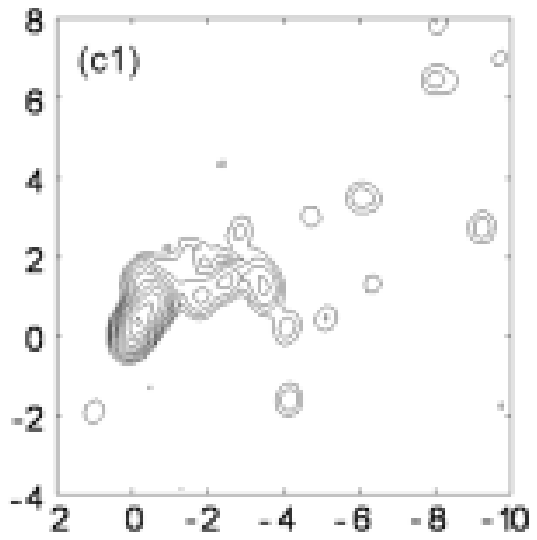}
\includegraphics[width = 40mm, height = 40mm]   {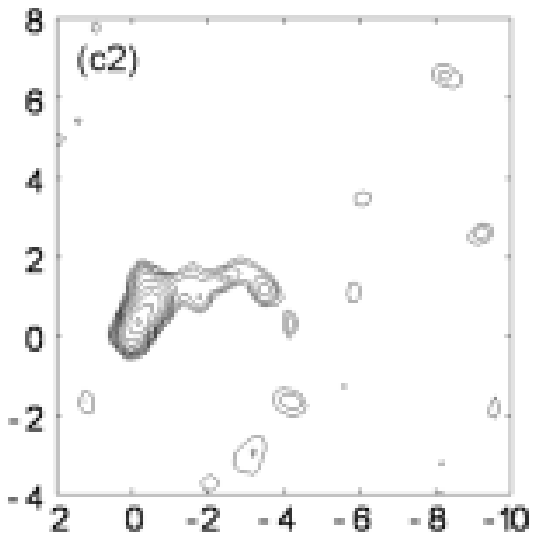}
\includegraphics[width = 40mm, height = 40mm]   {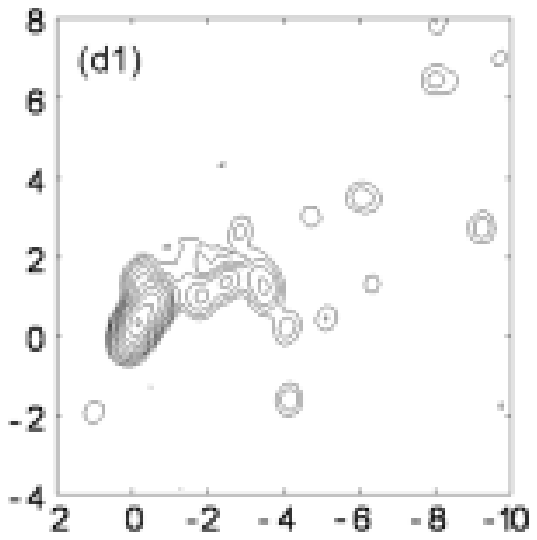}
\includegraphics[width = 40mm, height = 40mm]   {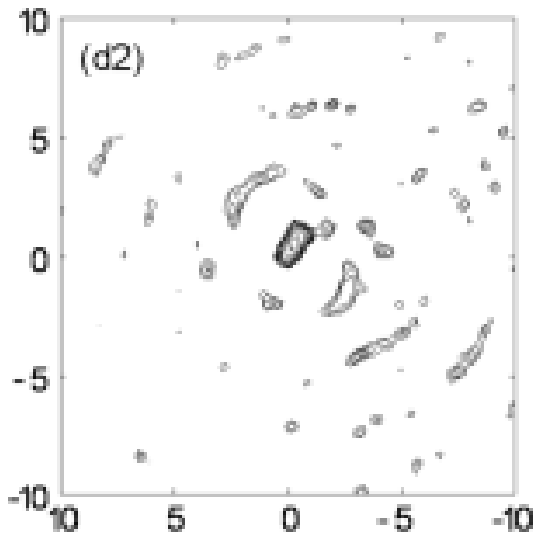}
\caption{Model (left column) and reconstructed (right column)
images at 5 (a), 8 (b), 15 (c) and 22~GHz (d). Relative right
ascension and declination are given in mas along the horizontal
and vertical axes, respectively. Contour levels: 0.2, 0.4, 0.8,
1.6, 3.2, 6.4, 12.8, 25.6, 51.2, 90\% of the peak flux density
(see Tables~1 and 2).} \label{fig:modsfs}
\end{center}
\end{figure}

Model maps of the source at frequencies of 5, 8, 15 and 22~GHz are
shown in Fig.~\ref{fig:modsfs}, left column. These maps were
obtained from the model map shown in Fig.~\ref{fig:mod13}, left,
according to the model spectral index distribution
(Fig.~\ref{fig:mod13}, right) and expression (\ref{eq:1}) for the
spectral dependence. Parameters of the model source intensity
maps, such as total flux density ($S_{\rm tot}$), peak flux
density ($S_{\rm peak}$) and entropy, are given in
Table~\ref{tbl:t1}. Comparison of the maps and their parameters
shows although not strong but still quite noticeable frequency
dependence.

The complex visibility functions, computed in accordance with the
Van Cittert-Cernike theorem, consists of 172 samples for each
``observation'' frequency. Each visibility value was aggravated
with additive random error to form data with a typical
signal-to-noise ratio of averaged and self-calibrated VLBI data of
about 10$^4$. It is necessary to emphasize that we do not consider
here the selfcalibration problems \citep{ccw90} and concentrate
only on removing spectral errors.

The main goals of the simulation fulfilled were: (i) to show
efficiency of the multi-frequency synthesis for improving
intensity images in case of small-element interferometers with
poor {\it uv}-plane filling; (ii) to illustrate the consequences
of multi-frequency synthesis without any spectral correction;
(iii) to estimate the possibility of reconstruction of spectral
index maps with a satisfactory quality.

We performed three tests. The first one was single-frequency
synthesis of the source images at 5, 8, 15 and 22~GHz. The second
one was multi(four)-frequency synthesis without any spectral
correction. And, finally, the third experiment was devoted
directly to multi-frequency synthesis with spectral correction, at
the reference frequency of 13.6~GHz.

The single-frequency maps are presented in Fig.~\ref{fig:modsfs},
right column. Their parameters are given in Table~\ref{tbl:t2}.
Analysis of the results obtained shows the following. The amount
of data proves to be too small, and {\it uv}-plane filling too
poor to obtain images with sufficient quality. Comparison of
Table~\ref{tbl:t1} and \ref{tbl:t2} allows us to judge about
distortions of the reconstructed images. As a criterium of the
reconstructed image quality, we chose the signal-to-noise ratio
(SNR) listed in the last column in Table~\ref{tbl:t2}, which was
calculated in the following way:
$$
{\rm SNR}=\frac{\sqrt{\sum_{k,l}I_{\rm
mod}^2(k,l)}}{\sqrt{\sum_{k,l}(I_{\rm mod}(k,l)-I_{\rm
rec}(k,l))^2}},
$$
where $I_{\rm mod}$ is the model map, $I_{\rm rec}$ is the
reconstructed map, $k,l=1,\ldots,N$, where $N$ is the linear size
of the map.

As it is seen from the single-frequency maps, at lower observation
frequencies, the lower resolution is ensured.  The lower-frequency
maps show a larger-scale structure, while the higher-frequency
maps show smaller-scale features. The highest accuracy of
reconstruction is achieved at frequencies of 8 and 15~GHz (${\rm
SNR}\sim16$). At the lowest frequency,  5~GHz, and the highest
frequency, 22~GHz, the obtained reconstruction quality was much
worse (SNR is about 8).

\begin{figure}
\begin{center}
\includegraphics[width = 80mm] {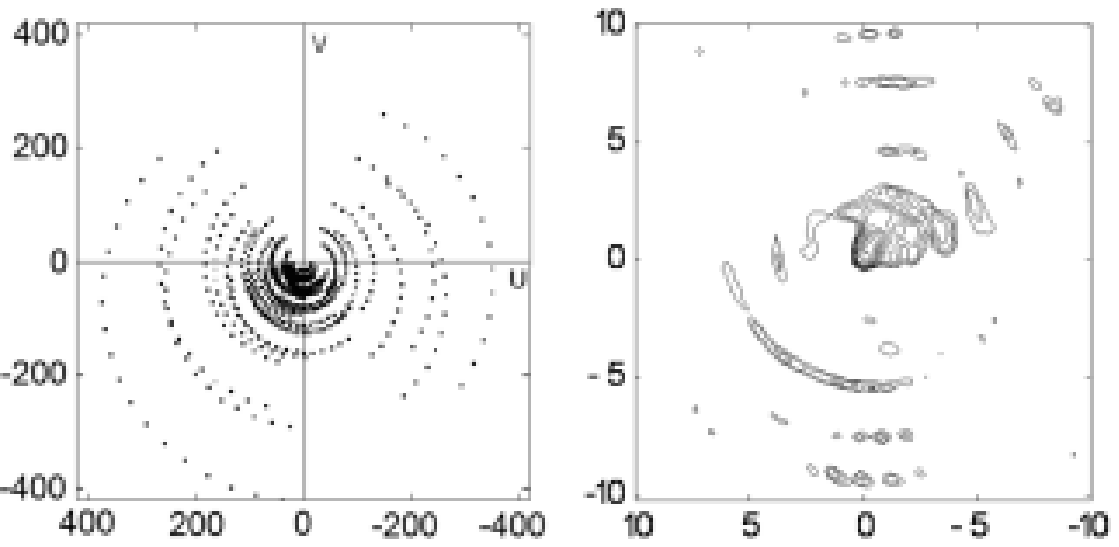}
\caption{Coverage of the {\it uv}-plane related to
multi(four)-frequency data (left) and the image reconstructed from
the four-frequency data ignoring any frequency dependence of the
source brightness distribution (right). The $u$- and $v$-axes are
in units of millions of wavelengths. Relative right ascension and
declination are given in mas along the horizontal and vertical
axes, respectively. Contour levels: 0.2, 0.4, 0.8, 1.6, 3.2, 6.4,
12.8, 25.6, 51.2, 90\% of the peak flux density of 0.194
Jy/pixel.} \label{fig:modnoncorr}
\end{center}
\end{figure}

The {\it uv}-plane corresponding to multi(four)-frequency
synthesis is presented in Fig.~\ref{fig:modnoncorr}, left. The
multi-frequency map obtained without any spectral corrections of
the frequency-dependent source brightness distribution is shown in
Fig.~\ref{fig:modnoncorr}, right. Parameters of the synthesized
map are given in Table~\ref{tbl:t3}. Comparing with the model map
(Fig.~\ref{fig:mod13}), we can see large image distortions which
are reflected in such map parameters as the total flux density,
entropy and SNR ($\sim$ 7). Large image distortions in the case of
relatively small values of spectral index distribution
(Fig.~\ref{fig:mod13}, right) can be explained by a wide frequency
range of the data.

\begin{figure}
\begin{center}
\includegraphics[width = 80mm] {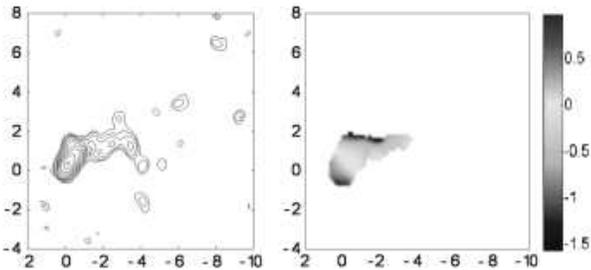}
\caption{Intensity image (left) and spectral index distribution
(right) reconstructed from the four-frequency data using the
GMEM-based algorithm with spectral correction (reference frequency
is 13.6~GHz). Relative right ascension and declination are given
in mas along the horizontal and vertical axes, respectively.
Contour levels: 0.2, 0.4, 0.8, 1.6, 3.2, 6.4, 12.8, 25.6, 51.2,
90\% of the peak flux density of 0.198 Jy/pixel.}
\label{fig:modmfs}
\end{center}
\end{figure}

The intensity map and spectral index image obtained using the
multi-frequency synthesis algorithm with spectral correction are
shown in Fig.~\ref{fig:modmfs}. Note that we tried the different
numbers of spectral terms in expansion (\ref{eq:2}). Here we
present the results related to utilization of three spectral
terms. Further increase in the number of spectral terms did not
improve our results substantially. Using fewer spectral terms
proved to be insufficient due to the wide frequency range of the
data. Having analyzed the results (Fig.~\ref{fig:modmfs} and
Table~\ref{tbl:t3}), we can conclude that taking into account the
frequency dependence of the source brightness distribution allowed
us to obtain both the source intensity map (${\rm SNR}\sim36$) and
spectral index distribution over the source with high accuracy.

Thus, in the source model with sufficiently complicated extended
structure, typical for AGN at milliarcsecond scales, with typical
spectral index values, we demonstrated the ability of the
MEM-based multi-frequency synthesis algorithm with spectral
correction for both  (i) improving intensity images and (ii)
obtaining spectral index maps of high quality. We emphasize that
the use of the MFS algorithm is especially effective in the case
of small-element interferometers with poor {\it uv}-plane filling.
We also showed the consequences of ignoring frequency dependence
of the source brightness distribution in multi-frequency synthesis
algorithm.

\begin{table}
\caption{Parameters of single-frequency reconstructed images.}
\label{tbl:t2}
\begin{center}
\begin{tabular}{ccccc}
\hline
Frequency         &$S_{\rm tot}$&$S_{\rm peak}$& Entropy &           SNR\\
 (GHz)            &         (Jy)&    (Jy/pixel)&         &              \\
\hline
\phantom{1}5.0& 5.16        & 0.168        & $-$18.3 &\phantom{1}7.4\\
\phantom{1}8.4& 4.71        & 0.187        & $-$16.1 &          16.8\\
          15.4& 4.54        & 0.197        & $-$15.5 &          16.5\\
          22.2& 4.41        & 0.189        & $-$15.7 &\phantom{1}8.4\\
\hline
\end{tabular}
\end{center}
\end{table}

\section{The problem of image alignment}
\label{s:align}

One of the most important sources of information about the
physical conditions in the radio-emitting regions of AGN is the
spectral index distribution over the source. The core region is
usually characterized by a large optical depth and an almost flat
or inverted spectrum, while the jets are optically thin with
respect to synchrotron radiation and have steeper spectra
\citep{crogab08, sulgab09, push05}.

\begin{table}
\caption{Parameters of four-frequency reconstructed images.}
\label{tbl:t3}
\begin{center}
\begin{tabular}{lcccc}
\hline
MFS experiment     &$S_{\rm tot}$&$S_{\rm peak}$&Entropy& SNR          \\
                   &(Jy)         &  (Jy/pixel)  &       &              \\
\hline
No spectral corr.  & 5.85        & 0.194        &$-$22.7&\phantom{1}7.2\\
With spectral corr.& 4.71        & 0.198        &$-$16.6&          36.2\\
\hline
\end{tabular}
\end{center}
\end{table}

The spectral index distribution over the source can be constructed
by various methods. The traditional method suggests: (i) formation
of images at two separate frequencies, $\nu_1$ and $\nu_2$, with
the solutions of the deconvolution problem (CLEAN or MEM) being
convolved with the same clean beam corresponding to the lower
observation frequency; (ii) calculation of the two-dimensional
spectral index distribution over the source from equation
(\ref{eq:1}). Obviously, this sequence of operations is legitimate
only when the positions of the VLBI cores of sources (not to be
confused with the physical core of the source that is undetectable
due to absorption effects) are frequency-independent.

The image reconstruction using the iterative selfcalibration
procedure is known \citep{thoms86} to lead to the loss of
information about the absolute position of the source on the sky:
during the phase self-calibration, the centroid of the object is
placed at the phase center of the map with coordinates $(0,0)$.
However, since most of the radio-loud AGN are characterized by a
dominant compact core \citep{kkl05,lister05,llk08}, the VLBI core
of the source coincides with the peak radio brightness of the
source in an overwhelming majority of cases.

\begin{figure}
\begin{center}
\includegraphics[width = 80mm, height = 40mm]   {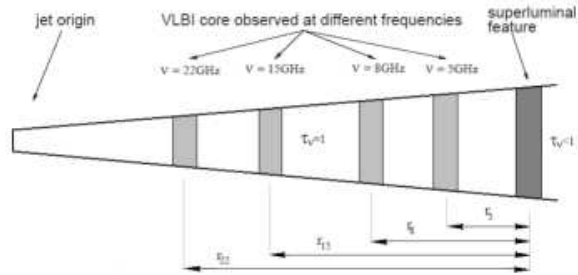}
\caption{A scheme illustrating the frequency-dependent position
shift of the VLBI core (modified from Kovalev et al. (2008a)).}
\label{fig:VLBIcore}
\end{center}
\end{figure}

Nevertheless, the standard theory of extragalactic radio sources
\citep{bk97} predicts a frequency-dependent VLBI core shift due to
opacity effects in the source's core region. Synchrotron
self-absorption takes place in an ultra-compact region near the
``central engine'' of AGN, the mechanism of which is most
efficient at low frequencies. As a result, the apparent origin of
the jet manifests itself farther from the physical core along the
jet axis at lower frequencies (Fig.~\ref{fig:VLBIcore}). This
theoretical prediction was confirmed by observations: the
frequency-dependent shift in the core position was measured for
several quasars by \cite{loban98}. In the literature, this
phenomenon is also actively debated from the viewpoint of the
accuracy of astrometric measurements \citep{charl02, bobol06,
kplz08}.

It thus follows that the multi-frequency data analysis must be
preceded by the alignment of images at different frequencies. This
can be achieved in three ways: (i) performing VLBI observations of
the objects under study together with reference sources; (ii)
finding the parameters of the shift of one image relative to the
other by aligning compact optically thin jet features, which are
not subject to absorption effects to the same extent as in the
source's core \citep{paragi00, kplz08}; (iii) finding the shift
parameters using a cross-correlation analysis \citep{crogab08}.
Being laborious from the viewpoint of performing observations and
their subsequent reduction, the first method gives no significant
advantage in determining the shift and its accuracy; therefore,
the second and/or third methods are used more often.

Recall that the alignment procedure implemented by shifting one
image relative to the other is equivalent to the phase correction
of the spectrum (or visibility function) of the image being
shifted relative to the fixed one. The need of precorrecting the
data for the source's visibility function at different frequencies
makes the direct use of the multi-frequency synthesis algorithm
described above problematic, because the frequency dependence of
the core shift is not known in advance. It can be determined by
forming the images at each frequency and determining the
corresponding shifts. As it was shown by \cite{kpls08},
\cite{sulgab09} and \cite{sok11}, the frequency dependence of the
VLBI core position is well fitted by a hyperbolic dependence of
the form $r\propto\nu^{-1}$. Thus, our multi-frequency synthesis
procedure can be used after allowance for the shifts in the
positions of the VLBI cores at different frequencies and their
coordinates relative to the phase center and applying the
corresponding frequency-dependent phase corrections to the
visibility function.

\section{Real data processing. Four-frequency imaging for 0954+658}
\label{s:proc}

Here we present the results of applying the developed
multi-frequency image synthesis algorithm to the  real VLBI data
of the extragalactic radio source 0954+658, a member of the
complete sample of BL~Lacertae objects~\citep{kuhr90}. Note that
0954+658 is also a member of the 1FGL catalog of $\gamma$-ray
bright sources detected by the Large Area Telescope onboard the
{\it Fermi} observatory and positionally associated with the
$\gamma$-ray source 1FGL J1000.1+6539~\citep{1FGL_AGN}. This
source is of our interest because it has a  typical parsec-scale
morphology that includes an optically thick VLBI core and a
one-sided optically thin jet, which is expected to manifest itself
in the spectral index distribution. General information on this
source is given in Table~\ref{tbl:t4}.

\begin{table}
\caption{General parameters of 0954+658.}
\label{tbl:t4}
\begin{center}
\begin{tabular}{cccccc}
\hline
Other name& Right    & Decli-  & Opt.& red-  & 1FGL \\
          &ascension &nation   & ID  & shift &      \\
          &(J2000)   &(J2000)  &     & ($z$) &      \\
\hline
J0958+6533& 9h 58m    & $+65\degr 33\arcmin$ & BL  & 0.367 & Y \\
          & 47.2451s & $54.818\arcsec$& Lac &      &  \\
\hline
\end{tabular}
\end{center}
\end{table}

The VLBA observations of 0954+658 were carried out in a
``snapshot'' mode in 1997 April (1997.26) simultaneously at four
frequencies: 5, 8, 15 and 22~GHz. The data were calibrated in the
NRAO AIPS package using standard procedures. The images were
formed within the framework of the Pulkovo ``VLBImager'' software
package based on a self-calibration algorithm \citep{cornfom99} in
combination with a GMEM-based deconvolution procedure.

\begin{figure}
\begin{center}
\includegraphics[width = 40mm, height = 40mm]   {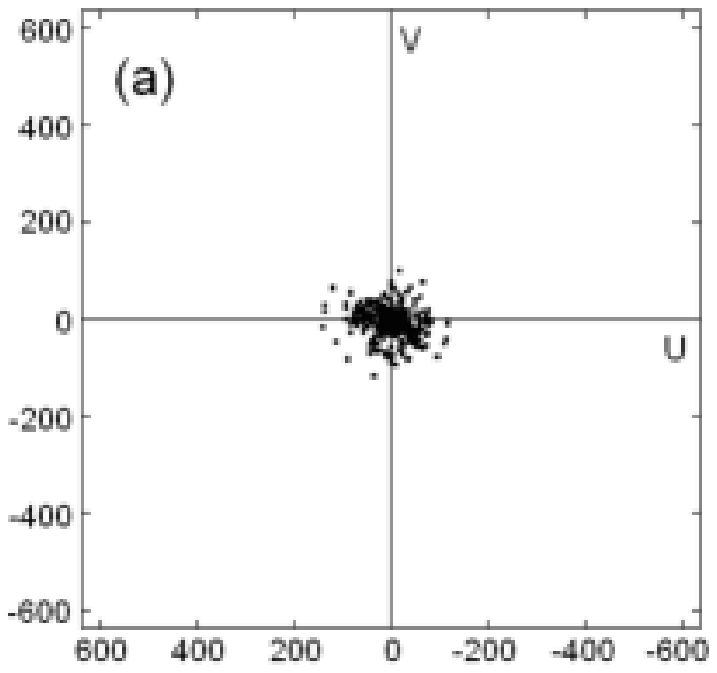}
\includegraphics[width = 40mm, height = 40mm]   {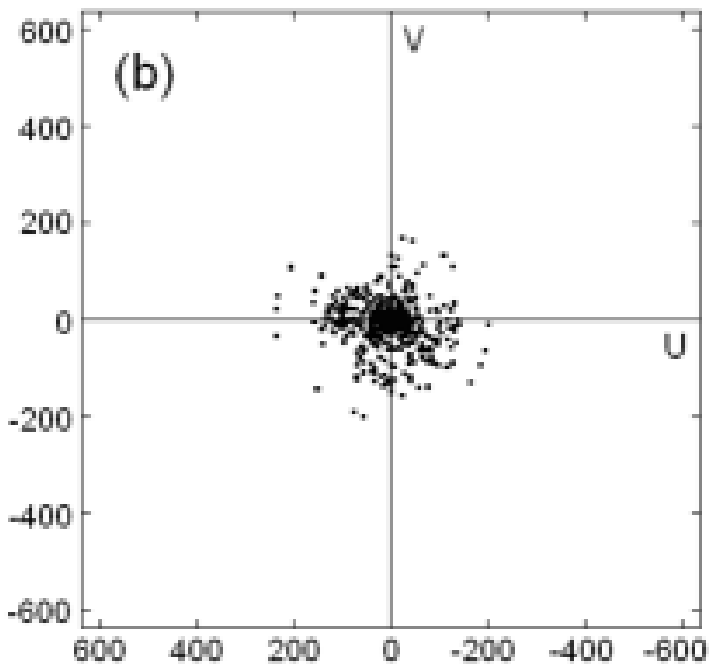}
\includegraphics[width = 40mm, height = 40mm]   {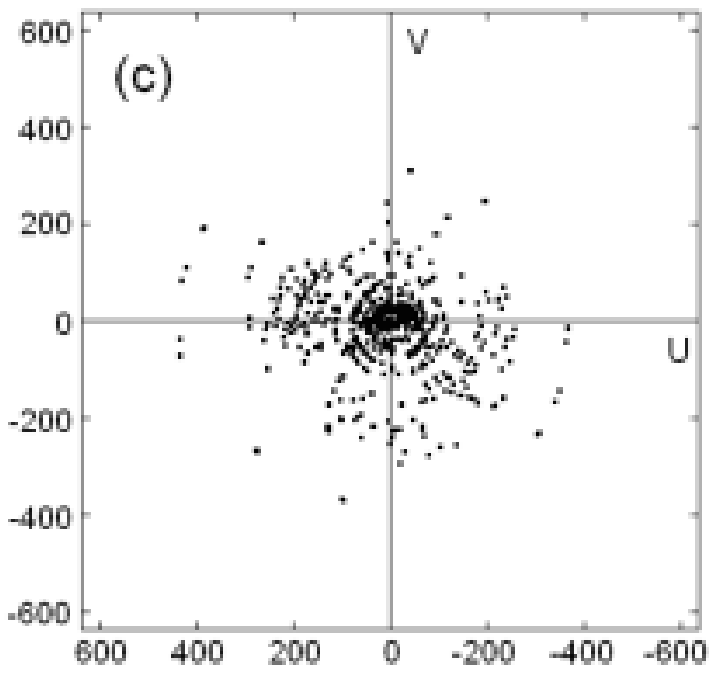}
\includegraphics[width = 40mm, height = 40mm]   {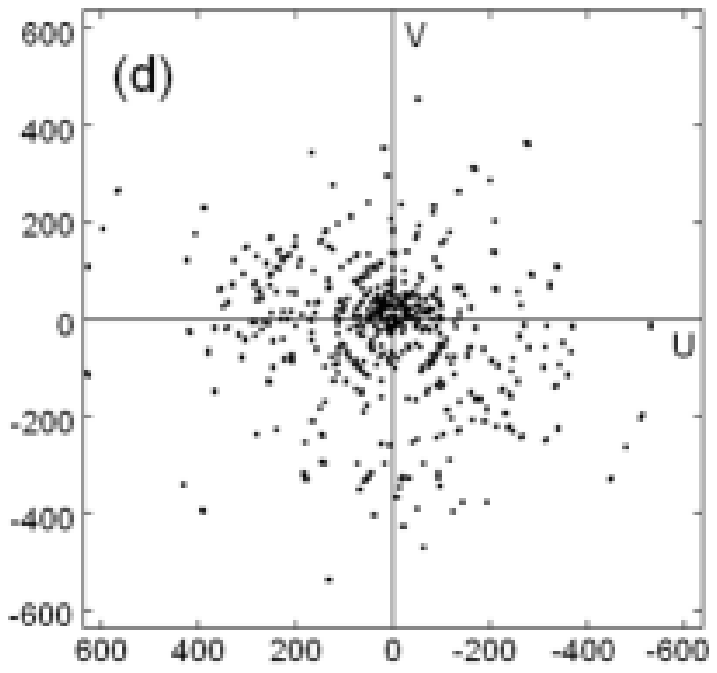}
\caption{Coverages of the {\it uv}-plane at 5 (a), 8 (b), 15 (c)
and 22 (d) GHz  for VLBA observations of 0954+658. The axes are in
units of millions of wavelengths.} \label{fig:VLBAuv}
\end{center}
\end{figure}

\begin{figure}
\begin{center}
\includegraphics[width = 40mm, height = 40mm]   {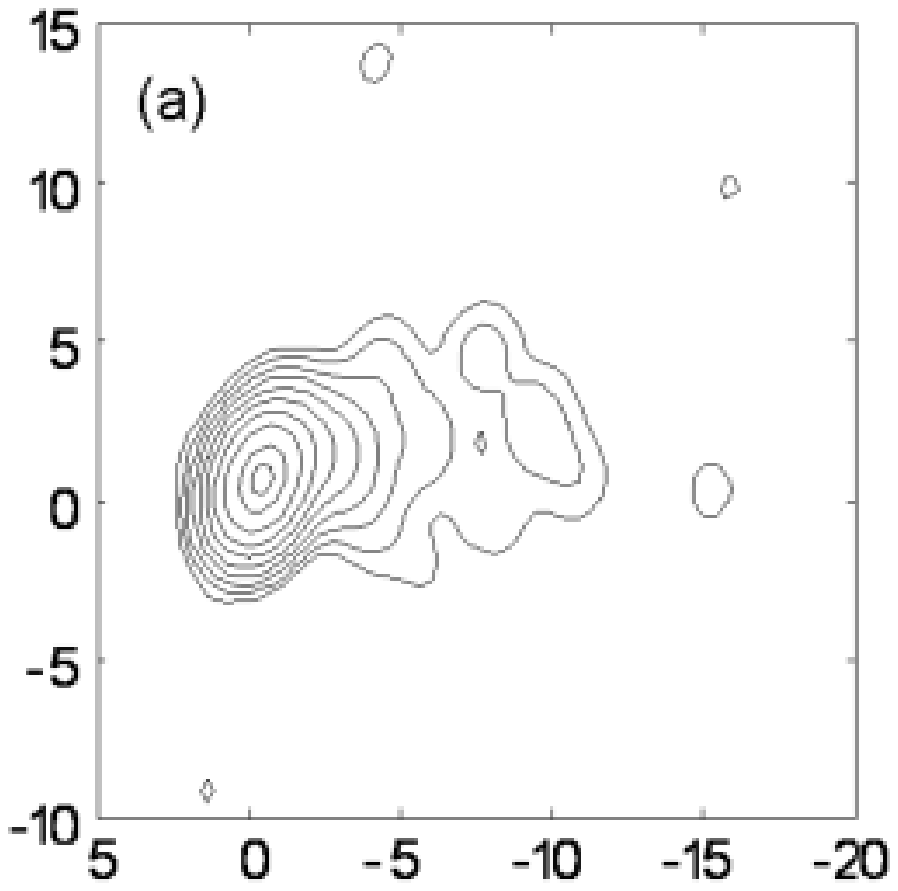}
\includegraphics[width = 40mm, height = 40mm]   {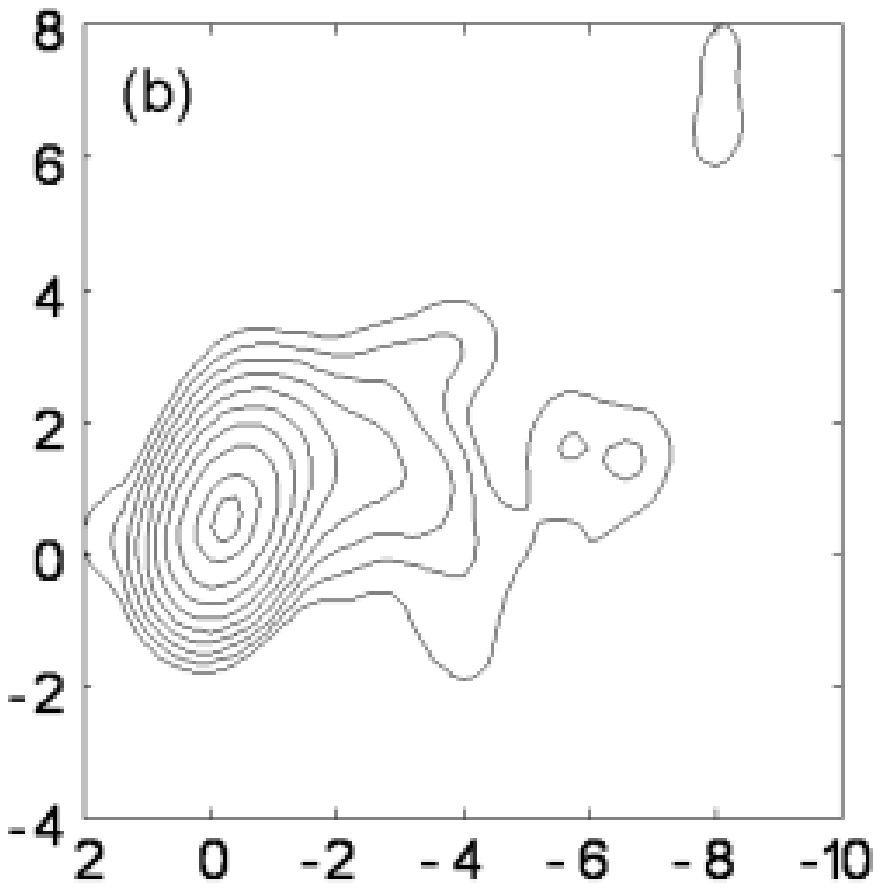}
\includegraphics[width = 40mm, height = 40mm]   {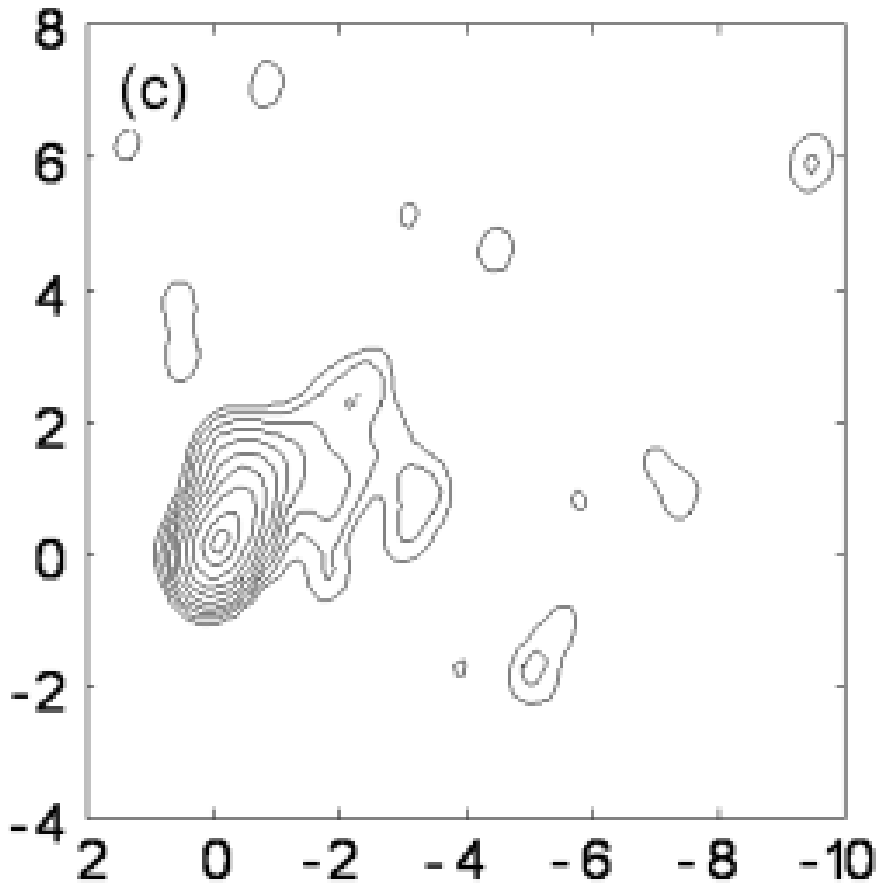}
\includegraphics[width = 40mm, height = 40mm]   {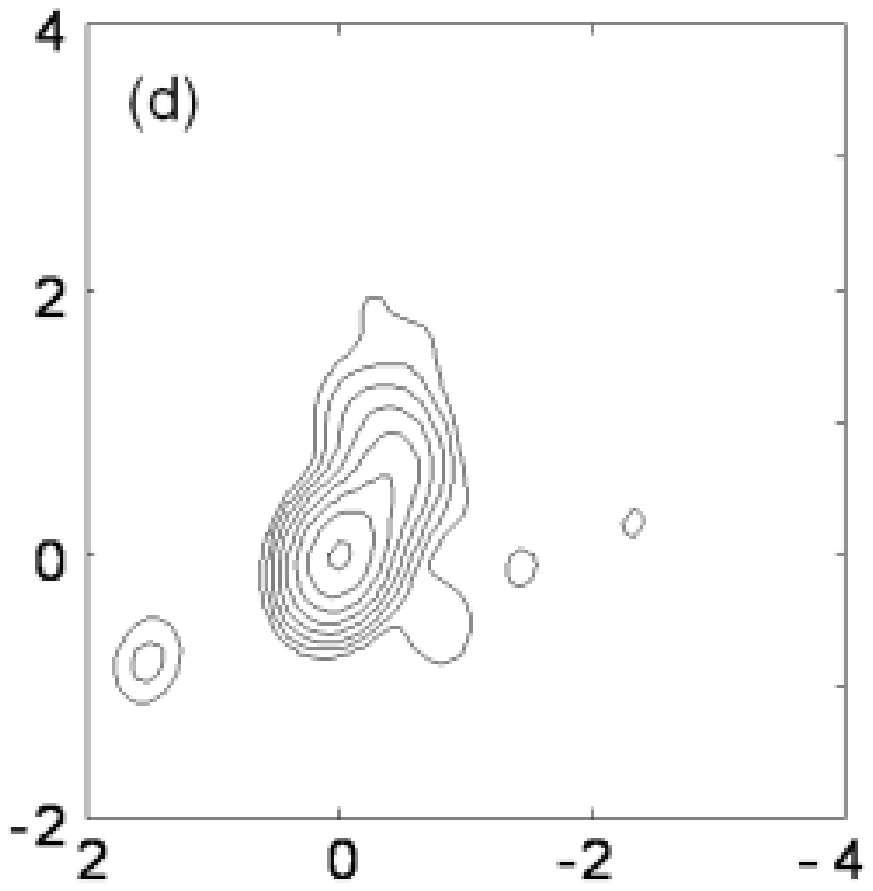}
\caption{Single-frequency maps of the source 0954+658
reconstructed from the data at 5 (a), 8 (b), 15 (c) and 22 (d)
GHz. Relative right ascension and declination are given in mas
along the horizontal and vertical axes, respectively. The lowest
contour levels are given in Table~\ref{tbl:t5} for each image, the
values of the succeeding levels are doubled.} \label{fig:VLBAsfs}
\end{center}
\end{figure}

The {\it uv}-plane  coverages  related to the observation
frequencies of 5, 8, 15 and 22~GHz are shown in
Fig.~\ref{fig:VLBAuv}. The MEM-based single-frequency maps are
shown in Fig.~\ref{fig:VLBAsfs}. Parameters of these maps,
obtained from the MEM-solutions by convolution with ``clean''
beams that determine the system's resolution at each observation
frequency, are given in Table~\ref{tbl:t5}.

\begin{table}
\caption{Parameters of single-frequency maps for 0954+658.}
\label{tbl:t5}
\begin{center}
\begin{tabular}{cccccc}
\hline
Freq-         &$S_\rmn{tot}$&$S_\rmn{peak}$&Beam               & Lowest      \\
uency         &      mJy    &  mJy/        &FWHM               &contour      \\
 GHz          &             &beam          &mas$\times$mas, PA & level, $\%$ \\
\hline
\phantom{1}5.0&   617       &   360        &2.50$\times$1.72, $-$17\fdg3&0.20 \\
\phantom{1}8.4&   496       &   311        &1.44$\times$1.05, $-$16\fdg2&0.10 \\
          15.4&   454       &   219        &0.81$\times$0.59, $-$16\fdg0&0.25 \\
          22.2&   310       &   166        &0.58$\times$0.43, $-$21\fdg6&0.70 \\
\hline
\end{tabular}
\end{center}
\end{table}

\begin{figure}
\begin{center}
\includegraphics[width = 80mm] {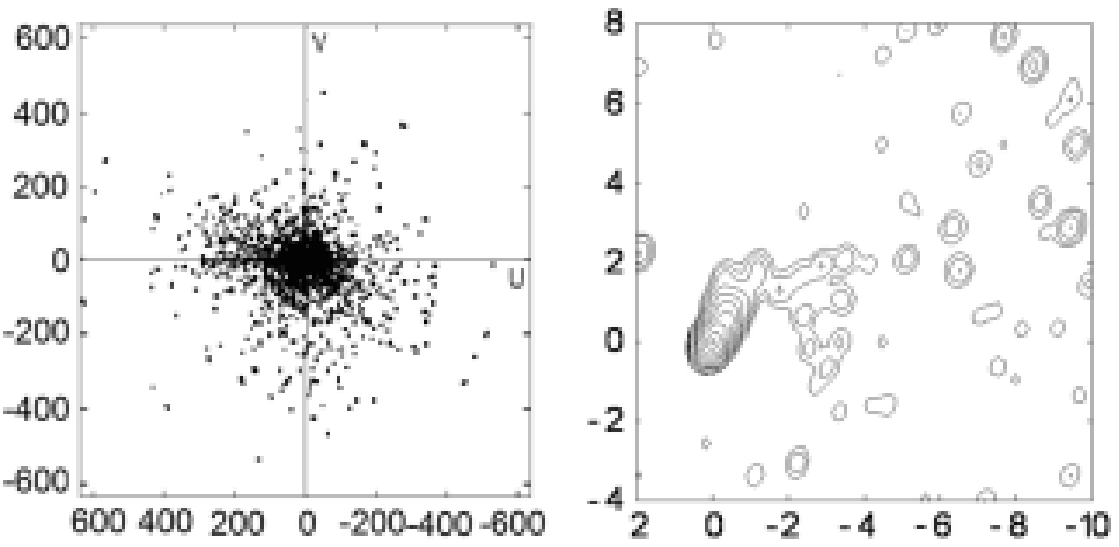}
\caption{Coverage of the {\it uv}-plane related to
multi(four)-frequency data (left) and the image reconstructed from
the four-frequency data ignoring any frequency dependence of the
source brightness distribution (right). The $u$- and $v$-axes are
in units of millions of wavelengths. Relative right ascension and
declination are given in mas along the horizontal and vertical
axes, respectively. Total VLBA flux density is 434 mJy. Beam FWHM
is 0.58$\times$0.43 mas, $\rmn{PA}=-$21\fdg6. Contour levels: 0.2,
0.4, 0.8, 1.6, 3.2, 6.4, 12.8, 25.6, 51.2, 90\% of the peak flux
density of 179 mJy/beam.}
 \label{fig:VLBAnoncorr}
\end{center}
\end{figure}

The parameters of the frequency-dependent image shift found by
aligning compact features of the optically thin jet are given in
Table~\ref{tbl:t6}. As expected, the direction of the shift
coincides with the inner jet orientation.  The {\it uv}-plane
related to the multi(four)-frequency data is shown in
Fig.~\ref{fig:VLBAnoncorr}, left.

\begin{table}
\caption{Frequency-dependent VLBI core position shifting
parameters with respect to the core position at 22~GHz.}
\label{tbl:t6}
\begin{center}
\begin{tabular}{ccc}
\hline
Frequency, GHz & $\Delta_r$, mas& Position Angle  \\
\hline
          15.4 & 0.18        & $-$29\fdg4         \\
\phantom{1}8.4 & 0.46        & $-$27\fdg1         \\
\phantom{1}5.0 & 0.73        & $-$29\fdg7         \\
\hline
\end{tabular}
\end{center}

\end{table}

\begin{figure}
\begin{center}
\includegraphics[width = 80mm] {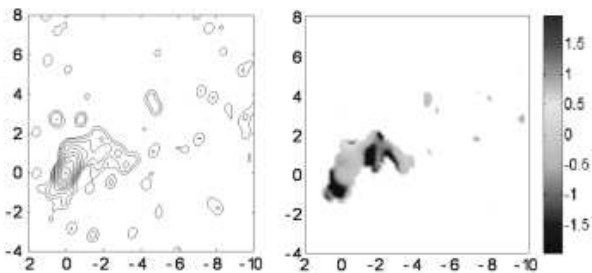}
\caption{Intensity (left) and spectral index (right) maps
reconstructed from the four-frequency data using the GMEM-based
algorithm with spectral correction, but without preliminary data
correction for frequency-dependent core position shift (reference
frequency is 13.6~GHz). Relative right ascension and declination
are given in mas along the horizontal and vertical axes,
respectively. Total VLBA flux density is 450 mJy. Beam FWHM is
0.58$\times$0.43 mas, $\rmn{PA}=-$21\fdg6. Contour levels: 0.2,
0.4, 0.8, 1.6, 3.2, 6.4, 12.8, 25.6, 51.2, 90\% of the peak flux
density  of 151 mJy/beam.} \label{fig:VLBAnoalign}
\end{center}
\end{figure}

First, we processed  the multi-frequency data ignoring dependence
of the source brightness distribution on  observation frequency.
The image obtained is shown in Fig.~\ref{fig:VLBAnoncorr}, right.
Then we applied our MFS algorithm with spectral correction to the
multi-frequency data, but we did not correct the data in
accordance with the frequency-dependent VLBI core position shift
found. Results of this multi-frequency synthesis obtained at
central reference frequency 13.6~GHz are shown in
Fig.~\ref{fig:VLBAnoalign}.

\begin{figure}
\begin{center}
\includegraphics[width = 80mm] {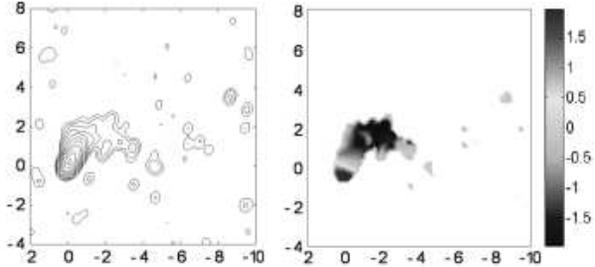}
\caption{Intensity (left) and spectral index (right) maps
reconstructed from the four-frequency data using the GMEM-based
algorithm with spectral correction and preliminary data correction
for frequency-dependent core position shift (reference frequency
is 13.6~GHz). Relative right ascension and declination are given
in mas along the horizontal and vertical axes, respectively. Total
VLBA flux density is 496 mJy. Beam FWHM is 0.58$\times$0.43 mas,
$\rmn{PA}=-$21\fdg6. Contour levels: 0.2, 0.4, 0.8, 1.6, 3.2, 6.4,
12.8, 25.6, 51.2, 90\% of the peak flux density of 189 mJy/beam.}
\label{fig:VLBAmfs}
\end{center}
\end{figure}

Finally, we applied our multi-frequency synthesis algorithm with
spectral correction to the observational data that were first
corrected according to the alignment parameters
(Table~\ref{tbl:t6}). The image obtained at reference frequency of
13.6~GHz and the corresponding two-dimensional spectral index
distribution are shown in Fig.~\ref{fig:VLBAmfs}.

The intensity images shown in Fig.~\ref{fig:VLBAnoncorr},
\ref{fig:VLBAnoalign} and \ref{fig:VLBAmfs} are obtained by
convolving solutions for $I_0$ with appropriate gaussian beam. The
spectral index images (Fig.~\ref{fig:VLBAnoalign} and
\ref{fig:VLBAmfs}) are formed by dividing two solutions for $I_1$
and $I_0$ in accordance with equation (\ref{eq:5}) after
convolving them with the same beam.

It is necessary to note that the main goal of this section is to
demonstrate the necessity of taking into account the
frequency-dependent image shift in order to correctly map the
spectral index distribution.

Let us analyze the results obtained.

The VLBI structure of the radio source 0954+658 consists of an
optically thick core (Fig.~\ref{fig:VLBAmfs}, right) and an
optically thin jet initially extending in the North-West direction
up to $\sim2$~mas from the VLBI core and then turning to the West.
As it is seen from single-frequency maps Fig.~\ref{fig:VLBAsfs}),
the apparent extent of the jet reaches about 12 mas and 2 mas at
the lowest and the highest frequencies, respectively. Total flux
density of the source varies from 0.6 up to 0.3~Jy at the lowest
and the highest frequencies, respectively. One can see that the
images obtained manifest large-scale structure features at lower
observation frequencies and small-scale ones at higher
frequencies.

Combining the data obtained at different observation frequencies
allows us, in general, to reconstruct both large- and small-scale
structures to a larger extent because of a better filled {\it
uv}-plane. That was proved, in particular, in Section \ref{s:test}
in model experiments.

\begin{figure}
\begin{center}
\includegraphics[width = 70mm]   {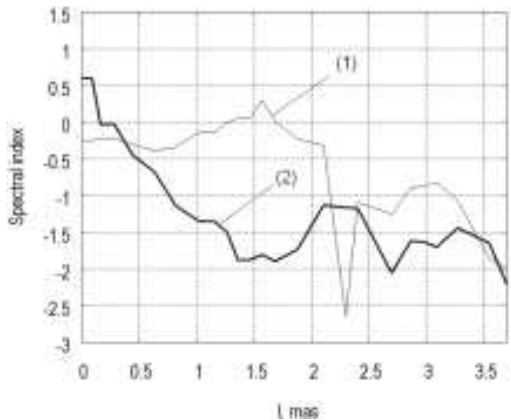}
\caption{Slices of the spectral index distributions along the jet
ridge line beginning from the phase center: (1) and (2) are
related to the spectral index maps presented in
Fig.~\ref{fig:VLBAnoalign} and Fig.~\ref{fig:VLBAmfs}
respectively; $l$ is the distance along the ridge line.}
\label{fig:VLBAcross}
\end{center}
\end{figure}

The intensity map (Fig.~\ref{fig:VLBAnoncorr}) synthesized
directly from the linearly combined multi-frequency data without
any spectral corrections shows large source structure distortions
and many artifacts. Multi-frequency synthesis with spectral
correction, fulfilled first without taking into account
frequency-dependent VLBI core position shift, shows a slight
improvement of the source intensity map, but it is still
distorted, especially in the South-West direction
(Fig.~\ref{fig:VLBAnoalign}). But the main consequence of image
misalignment is an incorrect reconstruction of the spectral index
distribution. In our case we can see that the spectral index map
obtained does not correspond to the physical meaning of an
optically thick core and an optically thin jet: there are negative
spectral indices in the core region and segments with positive
spectral indices in the jet region. From Fig.~\ref{fig:VLBAmfs},
we realize that only allowance for the real shift found by
aligning features of the optically thin jet yielded the proper
result. As it is seen, we managed to reconstruct a more extended
jet structure (up to 8 mas along the RA axis) than in the case of
using only the high-frequency data but with the same high angular
resolution. The spectral index map adequately reflects the
physical characteristics of the regions of the optically thick
compact VLBI core and the optically thin extended jet. We see a
fairly regular structure with smooth transitions between segments
of different intensities along the entire source. For comparison,
one-dimensional slices of the spectral index distributions (see
Fig.~\ref{fig:VLBAnoalign}, left and Fig.~\ref{fig:VLBAmfs}, left)
obtained along the jet ridge line are shown in
Fig.~\ref{fig:VLBAcross}, from which we can graphically evaluate
undesirable consequences of ignoring the frequency-dependent VLBI
core position shift in the MFS algorithm.

\section{Conclusions}
\label{s:concl}

We developed and tested an efficient multi-frequency image
synthesis algorithm with the correction for the frequency
dependence of the radio brightness of a source.  The algorithm is
based on the generalized maximum entropy method; it allows one to
take into account the spectral terms of any order and to map both
a total intensity image and spectral index distribution, which is
of great importance in investigating the physical characteristics
of AGN.

The advantage of the proposed multi-frequency synthesis algorithm
is that the spectral terms of any order can be easily taken into
account in the entropic functional being minimized. This allows
for the spectral correction of images to be made both in a wide
frequency range and for large spectral indices.

We have shown how important the allowance for the
frequency-dependent image shift is in applying the multi-frequency
synthesis algorithm. Our conclusions are based on the results of
processing the multi-frequency VLBA data for the BL Lac object
0954+658 with a fairly complex extended jet structure, which also
manifests itself in the spectral index distribution over the
source.

Analysis of the results obtained shows that multi-frequency
synthesis is an efficient method for improving the mapping
quality; low-frequency data allow the extended structure of a
source to be reconstructed more completely, while high-frequency
data allow a high spatial resolution to be achieved. It should
also be emphasized that the spectral index distribution can be
mapped with a high quality as well.

\section*{ACKNOWLEDGMENTS}
The authors are grateful to the referee for critical remarks,
which helped to improve the paper. This work was supported by the
``Origin and Evolution of Stars and Galaxies'' Program of the
Presidium of the Russian Academy of Sciences and the Program of
State Support for Leading Scientific Schools of the Russian
Federation (grant NSh-3645.2010.2 ``Multi-wavelength Astrophysical
Research''). The VLBA is a facility of the National Science
Foundation operated by the National Radio Astronomy Observatory
under cooperative agreement with Associated Universities, Inc. The
authors are thankful to Vladimir Kouprianov for his assistance in
preparing the text of the manuscript.

{
}

\begin{thebibliography}{99}

\bibitem[\protect\citeauthoryear{Abdo et al.}{2010}]{1FGL_AGN} Abdo A.~A., Ackermann M.,
Ajello M., Allafort A., Antolini E., Atwood  W.~B., Axelsson M.,
Baldini, L., Ballet J., Barbiellini G.,  et al., 2010, ApJ, 715,
429
\bibitem[\protect\citeauthoryear{Ables}{1974}]{abl74} Ables J. G., 1974, AASS, 15, 383
\bibitem[\protect\citeauthoryear{Bajkova}{1992}]{bajk92} Bajkova A.~T., 1992,
Astron. Astrophys. Trans., 1, 313
\bibitem[\protect\citeauthoryear{Bajkova}{1993}]{bajk93} Bajkova A.~T., 1993, Reports of IAA RAS, N 58 (in Russian)
\bibitem[\protect\citeauthoryear{Bajkova}{2005}]{bajk05} Bajkova A.~T., 2005, Astron. Rep., 49,
947
\bibitem[\protect\citeauthoryear{Bajkova}{2007}]{bajk07} Bajkova A.~T., 2007, Astron. Rep., 51,
891
\bibitem[\protect\citeauthoryear{Bajkova}{2008}]{bajk08} Bajkova A.~T., 2008, Astron.
Rep., 52, 951
\bibitem[\protect\citeauthoryear{Blandford \& K\"{o}nigl}{1979}]{bk97} Blandford R.~ D., K\"{o}nigl A., 1997,
APJ, 232, 34
\bibitem[\protect\citeauthoryear{Boboltz}{2006}]{bobol06} Boboltz
D.~A.,2006, {\it IERS Technical Note 34, Intern.l Celest.
Reference System and Frame}, Verlag des Bundesamtes f\"{u}ur
Kartographie and Geod$\ddot{a}$sie, Frankfurt-am-Main
\bibitem[\protect\citeauthoryear{Charlot}{2002}]{charl02} Charlot P., 2002, {\it Proc. of the Intern. VLBI Service for
Geodesy and Astrometry 2000, General Meeting},
NASA/CP-2002-210002, p. 233
\bibitem[\protect\citeauthoryear{Conway}{1991}]{conw91} Conway J.~E., 1991, ASP Conf. Ser., 19, 171
\bibitem[\protect\citeauthoryear{Conway, Cornwell \& Wilkinson}{1990}]{ccw90} Conway J.~E., Cornwell T. J., Wilkinson P. N., 1990,
MNRAS, 246, 490
\bibitem[\protect\citeauthoryear{Cornwell, Braun \& Briggs}{1999}]{cbb99} Cornwell T. J., Braun R., Briggs D.
S., 1999, {\it Synthesis imaging in Radio Astronomy II},  ASP
Conf. Ser., 180, 151
\bibitem[\protect\citeauthoryear{Cornwell \& Evans}{1985}]{coreva85} Cornwell T. J., K. F. Evans, 1985,
A\&A, 143, 77
\bibitem[\protect\citeauthoryear{Cornwell \& Fomalont}{1999}]{cornfom99} Cornwell, T. \& Fomalont, E.B. 1999, {\it Synthesis imaging in Radio Astronomy II},  ASP
Conf. Ser., 180, 187
\bibitem[\protect\citeauthoryear{Croke \& Gabuzda}{2008}]{crogab08} Croke S.~M., Gabuzda D.
C., 2008, MNRAS, 386, 619
\bibitem[\protect\citeauthoryear{Frieden}{1972}]{fried72} Frieden B.~R., 1972,
JOSA, 72, 511
\bibitem[\protect\citeauthoryear{Frieden \& Bajkova}{1994}]{frbajk94} Frieden B.~R., Bajkova A.
T.,1994,  Appl. Opt., 33, 219
\bibitem[\protect\citeauthoryear{H\"{o}gbom}{1974}]{hogb74} H\"{o}gbom
J.~A.,1974, AASS, 15, 417
\bibitem[\protect\citeauthoryear{Kardashev}{1997}]{kard97} Kardashev N.~S., 1997,
 Exp. Astron., 7, 329
\bibitem[\protect\citeauthoryear{Kovalev et al.}{2005}]{kkl05} Kovalev Y.~Y., Kellermann K. I.,
 Lister M. L., Homan  D. C., Vermeulen R. C., Cohen M. H.,
 Ros E., Kadler M., Lobanov A. P., J. A. Zensus, et al., 2005, AJ, 130, 2473
\bibitem[\protect\citeauthoryear{Kovalev et al.}{2008a}]{kplz08} Kovalev Y.~Y., Lobanov A. P.,
Pushkarev A. B., Zensus J. A., 2008, A\&A, 483, 759
\bibitem[\protect\citeauthoryear{Kovalev et al.}{2008b}]{kpls08} Kovalev Y.~Y., Pushkarev A. B.,
Lobanov A. P., K. V. Sokolovsky, 2008, {\it Proc. of the 9th Eur.
VLBI Network Symp. PoS}, p. 7
\bibitem[\protect\citeauthoryear{K\"{u}hr \& Schmidt}{1990}]{kuhr90} K\"uhr H., Schmidt G., 1990, AJ, 90, 1
\bibitem[\protect\citeauthoryear{Lee et al.}{2008}]{llk08} Lee S.-S., Lobanov A. P.,
 Krichbaum T. P.,  Witzel1 A., Zensus A., Bremer M.,
Greve A., Grewing M., 2008, AJ, 136, 159
\bibitem[\protect\citeauthoryear{Likhachev, Ladynin \&
Guirin}{2006}]{llg06} Likhachev S. F., Ladynin V. A., Guirin I.
A., 2006, Radioph. \& Quant. Electr., 49, 499
\bibitem[\protect\citeauthoryear{Lister \& Homan}{2005}]{lister05} Lister M.~L., Homan D.~H., 2005, AJ, 130, 1389
\bibitem[\protect\citeauthoryear{Lobanov}{1998}]{loban98} Lobanov A.~P., 1998, A\&A, 330, 79
\bibitem[\protect\citeauthoryear{Narayan \& Nityananda}{1986}]{narnit86} Narayan R., Nityananda R., 1986,
Ann. Rev. A\&A, 24, 127
\bibitem[\protect\citeauthoryear{Oppenheim \& Schafer}{1999}]{opp99} Oppenheim A. V., Schafer R. W., 1986, {\it Discrete-Time Signal Processing},
Prentice  Hall, NJ
\bibitem[\protect\citeauthoryear{Paragi, Fejes \& Frey}{2000}]{paragi00} Paragi Z., Fejes I., Frey
S., 2000, {\it Proc. of the Intern. VLBI Service for Geodesy and
Astrometry 2000, General Meeting}, NASA/CP-2000-209893, p. 342.
\bibitem[\protect\citeauthoryear{Pushkarev et al.}{2005}]{push05} Pushkarev A.~B., Gabuzda D. C.,
Vetukhnovskaya Yu. N., Yakimov V. E., 2005, MNRAS, 356, 859
\bibitem[\protect\citeauthoryear{Rastorgueva et al.}{2011}]{rast11} Rastorgueva E.~A.,
Wiik K. J., Bajkova A. T., Valtaoja E., Takalo L. O.,
Vetukhnovskaya Yu. N., Mahmud M., 2011, A\&A, 529, A2
\bibitem[\protect\citeauthoryear{Rau et al.}{2009}]{rau09} Rau U.,
Bhatnagar S., Voronkov M. A., Cornwell T. J., 2009, Proc. IEEE,
97, 1471
\bibitem[\protect\citeauthoryear{Rau}{2010}]{rau10} Rau U., 2010, {\it Parameterized Deconvolution
for Wide-band Radio Synthesis Imaging}, PhD thesis, New Mexico
Institute of Mining and Technology
\bibitem[\protect\citeauthoryear{Sault \& Wieringa}{1994}]{sault94} Sault R.~J., Wieringa M. H., 1994,
  AASS, 108, 585
\bibitem[\protect\citeauthoryear{Sault \& Conway}{1999}]{sault99} Sault R.~J., Conway J. E.,
1999, ASP Conf. Ser., 180, 419
\bibitem[\protect\citeauthoryear{Skilling \& Bryan}{1984}]{skil84} Skilling J., Bryan R.
K., 1984, MNRAS, 211, 111
\bibitem[\protect\citeauthoryear{Sokolovsky et al.}{2011}]{sok11} Sokolovsky
K. V., Kovalev Y. Y., Pushkarev A. B., Lobanov A. P., 2011, A\&A,
submitted
\bibitem[\protect\citeauthoryear{O'Sullivan \& Gabuzda}{2009}]{sulgab09} O'Sullivan, Gabuzda D. C.,
2009, MNRAS, 400, 26
\bibitem[\protect\citeauthoryear{Thompson, Moran \& Swenson}{2001}]{thoms86} Thompson A.~R., Moran J. M., Swenson G. W., 2001, {\it Interferometry and Synthesis in Radio
Astronomy}, Wiley-Intersci. Pub., New York
\bibitem[\protect\citeauthoryear{Wernecke \& D$^\prime$Addario}{1976}]{wern76}  Wernecke S. J., D$^\prime$Addario L. R., 1976, IEEE-C, 26, 351

\end{thebibliography}
\end{document}